\documentclass[aps,prb,twocolumn,groupedaddress]{revtex4}

\newcommand{\pp}{\mathcal{P}}
\newcommand{\fsize}{0.45\textwidth}

\usepackage{amssymb,amsmath}
\usepackage{graphicx}

\begin{document}

\title{Symmetry in a Perturbed Optical System}

\author{Jason H. Steffen}
\email[]{jsteffen@astro.washington.edu}
\affiliation{University of Washington, Department of Physics}
\date{\today}

\begin{abstract}
This paper investigates the effects that perturbations to an optical system, such as translations or rotations of the optical elements, have on the final location where a light ray strikes a detector.  Symmetry arguments are employed to give selection rules for the combinations of those perturbations that can affect the light ray.  It is shown that a ``Transverse Parity'' can be assigned to each type of perturbation and that only combinations with an overall odd parity will produce an effect.  Some examples of how these results can be applied to undergraduate classroom or laboratory courses are given.
\end{abstract}

\maketitle

\section{Introduction}

The study of physics is replete with the use of symmetry to explain physical phenomena.  The conservation of energy, momentum, and electric charge stem from symmetries as do the selection rules for allowed quantum-mechanical interactions or multipole moments of a force field or mass distribution.  Indeed, in the undergraduate curriculum one finds symmetries and their application in nuclear and particle physics, electromagnetism, quantum mechanics, and statistical mechanics to name a few.  It comes as no surprise that symmetries play a role in optics as well.  Indeed, once one moves beyond the nominal placement of the elements of an optical system and studies, instead, the effects of perturbations to such systems, the effects of symmetry in optics become more subtle and more rich.

This work investigates some roles that symmetry plays in an optical system that has been perturbed from its nominal configuration.  This investigation reveals straightforward relations between the perturbations (translations or rotations) of the optical elements that can combine to produce a displacement of the beam spot or a rotation of the incident angle of the beam as it impinges upon a screen or detector.  A straightforwared symmetry argument gives the selection rules for the allowed couplings among the perturbations.

The advantage of this development for the undergraduate is that, unlike most treatments of symmetry (e.g. quantum mechanics), an optical system is easily manipulated and the consequence of changes to the system are immediately apparent.  The relative transparancy of the effects of changes to an optical system allows the student to focus on understanding the underlying physical principles instead of the proper use of a given experimental apparatus.  With a few basic optical elements and a well designed set of exercizes, a physics student can study one of the most important aspects of modern physical theory and appreciate the power of using symmetries to describe physical scenarios.  The materials presented here are the aspects of a more general study that are readily incorporated into a standard undergraduate lecture or laboratory course.

The original motivation for this work was to examine the systematic effects that are introduced into an experiment due to imperfections in the placement and alignment of the optical elements in several, nominally-identical optical subsystems.  In particular, an array of eight optical systems was used to determine the parameters of motion of a torsion pendulum undergoing large-amplitude torsional oscillations.  Since the experiment required nanosecond timing precision, precise placement of the optical elements was important as well as understanding the leading systematic effects.  The results of that investigation are the basis for this work.

This paper will proceed by first introducing the approach that was used for developing the optical system model along with the assumptions that are made about the optical elements themselves.  Some short examples are provided to show the nature of the model and the results it bears.  Next, selection rules for the allowed couplings of optical element perturbations are demonstrated and derived via the introduction of a ``Transverse Parity'' that is assigned to the various ways that an element can be perturbed.  Finally, attention is given to the application of this work to the undergraduate classroom or laboratory.

\section{Model of the Optical System}

For this study, the optical system is modeled using a quasi object-oriented approach where the position and orientation information of a given lens, mirror, ray, etc. is encapsulated in a lens, mirror, or ray object.  For example, a ray object consists of a point of origin and a direction vector, a lens object consists of a focal length, a center point, and a normal vector.  The effect of a ray refracting through a lens is found using a function that takes a given ray and lens, then returns the refracted ray using analytic geometry.

This method is different from the standard geometrical or two-dimensional matrix formulation that is presented in most undergraduate optics texts~\cite{serway,hecht} and from more general three dimensional matrix approaches~\cite{almeida}.  This approach was chosen because it facilitates the three-dimensional manipulation of the optical elements and allows for a more straightforward interpretation of the analytic results that it produces.  Most of the computation was done with the symbol-manipulating software Mathematica.

The following assumptions are made throughout this work: all of the optical elements are perfect, thin, are sufficiently large that displacements and rotations of the elements still allow the beam to impinge upon them, and that refraction is not wavelength dependent.  It is also convenient to assume that the deviations of the optical elements from their nominal position are small so that they can be treated perturbatively~\footnote{When only displacements are considered, the presented formulas are exact---there are no higher order terms.}.  That is, rotations are small and displacements are small compared to the appropriate focal lengths.  Finally, only a single, central ray of the beam is considered in this investigation.  Some comments concerning an extended beam are made near the end of the paper.

\subsection{Example: Linear Displacement I}

\begin{figure}[!ht]
\begin{center}
\includegraphics[width=\fsize]{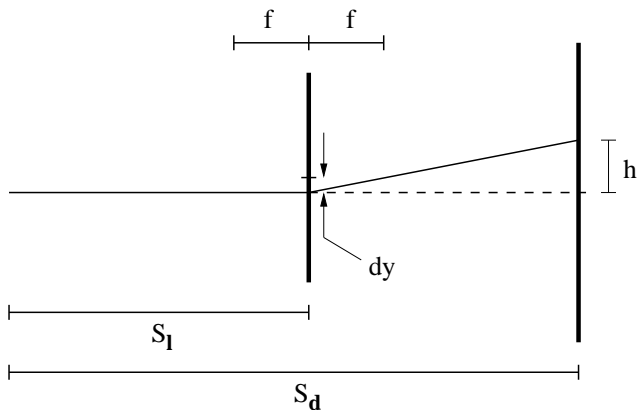}
\caption{A light ray refracting through a lens of focal length $f$ which is displaced vertically by an amount $dy$ from its nominal position.  This perturbation causes the final location of the beam to be displaced vertically by an amount $h$.}
\label{fig1}
\end{center}
\end{figure}

The first example to consider is that of a single lens and detector nominally placed along the longitudinal (or ``optic'') axis.  If the lens is displaced vertically a distance $dy$, as shown in figure \ref{fig1}, then the location where the beam strikes the detector is also displaced vertically by an amount
\begin{equation}
\label{single}
h = \frac{s_d-s_l}{f}dy
\end{equation}
where $f$ is the focal length of the lens, $s_l$ is the longitudinal position of the lens, and $s_d$ is the longitudinal position of the detector.  Equation (\ref{single}) shows that the vertical displacement of the beam spot is linear in the displacement of the lens and also linear in the separation of the lens and detector---the ``lever arm''.

\subsection{Example: Linear Displacement II}

\begin{figure}[!ht]
\begin{center}
\includegraphics[width=\fsize]{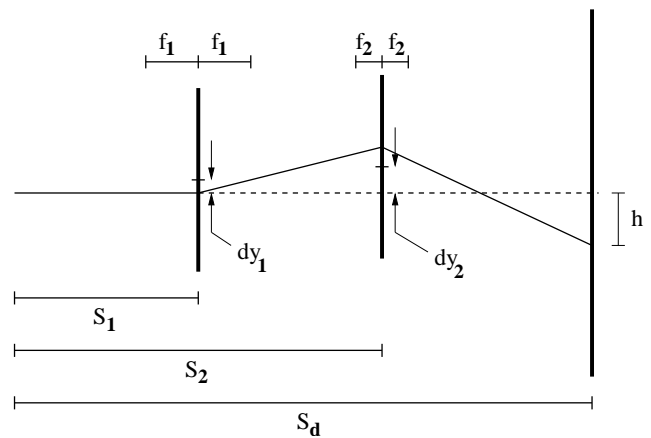}
\caption{A light ray refracting through two lenses.  One, with focal length $f_1$ is displaced vertically by an amount $dy_1$.  The other, with focal length $f_2$ is displaced by an amount $dy_2$.  The vertical displacement of the beam at the screen is given by $h$.}
\label{fig2}
\end{center}
\end{figure}

Now, suppose that two lenses are used in the optical system, again aligned nominally along the longitudinal axis.  If the two lenses are displaced vertically by amounts $dy_1$ and $dy_2$, as shown in figure \ref{fig2}, then the resulting vertical displacement of the beam spot is given by
\begin{equation}
\begin{split}
\label{double}
h=& \left( \left( \frac{s_d-s_1}{f_1}\right) - \left( \frac{s_2-s_1}{f_1} \right) \left( \frac{s_d-s_2}{f_2} \right) \right) dy_1 \\
&+ \left( \frac{s_d-s_2}{f_2} \right) dy_2
\end{split}
\end{equation}
where $s_1$, $s_2$, and $s_d$ are the positions of the first lens, the second lens, and the detector respectively and $f_1$ and $f_2$ are the focal lengths of the lenses.

This example shows that the displacement of the beam spot grows linearly with the vertical displacements of the lenses, as expected.  It also shows that there are two competing effects that depend upon the separation of the optical elements.  For small separations of the optical elements the spot height grows linearly with the nominal placements of the lenses.  For large separations of the elements, the height falls quadratically.

A more interesting feature of equation (\ref{double}) is that there is no coupling between the displacements $dy_1$ and $dy_2$.  Generally, one might expect that cross terms of the form $dy_1dy_2$ might exist, yet they do not.  If a third lens is included, then there are three terms in the resulting height of the beam spot.  Each term is linear in the displacement of a particular lens, and again, no cross terms exist.  This fact holds true for any displacements that lie in the plane that is transverse to the incoming beam.  This trend, however, does not apply for longitudinal displacements.  It can be seen from equation (\ref{single}) that a small displacement of the lens in the longitudinal direction ($s_l \rightarrow s_l + dz$) would contribute a quadratic term of the form $dy dz/f$.

\subsection{Example: Rotation}

\begin{figure}[!ht]
\begin{center}
\includegraphics[width=\fsize]{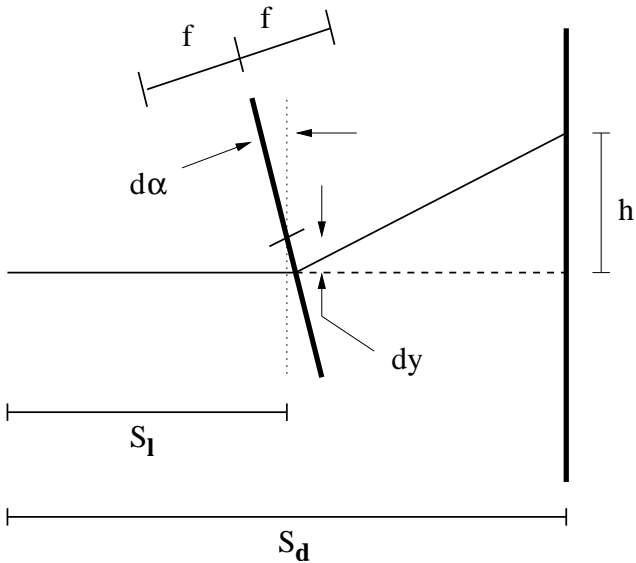}
\caption{A light ray is refracted by a lens of focal length $f$ which is displaced vertically a distance $dy$ and rotated by an angle $d\alpha$ about the $x$-axis.  The resulting displacement of the beam on the detector is given by $h$.}
\label{fig3}
\end{center}
\end{figure}

In the model used for this paper, each rotation of an optical element is about a body centered axis that remains parallel to the corresponding lab-frame axis.  Thus, the origin of the rotation axes move with the lens, but the orientation of the axes do not.  To see the effects of rotation, consider the case of a single lens displaced vertically a distance $dy$ and rotated about the $x$ axis by an angle $d\alpha$ as shown in figure \ref{fig3}.  This example, to third order in the perturbations, gives a vertical displacement of the beam on the detector
\begin{equation}
\begin{split}
\label{rotated}
h =& \left( \frac{s_d-s_l}{f} \right) dy + \left( \frac{f-(s_d-s_l)}{f^2}\right)dy^2 d\alpha \\
&- \frac{1}{2}\left( \frac{s_d-s_l}{f} \right)dy d\alpha^2.
\end{split}
\end{equation}
Unlike the previous examples where all the perturbations were displacements, here there are cross terms between the transverse displacements and the rotations of the lens about the transverse axes.  Regardless of the number of elements in the optical system, each term in equations like (\ref{rotated}) will contain an odd number of transverse perturbative factors (i.e. factors that correspond to perturbations in the transverse directions).  Similar to the previous examples, if there is a rotation about the longitudinal axis, then some couplings will contain an even number of perturbative factors.  However, the terms that involve longitudinal perturbations still have an odd number of transverse perturbations.  The selection rules for terms that ultimately contribute to the displacement of the beam spot are addressed in the next section.

\section{Transverse Parity}

The examples presented in the last section show that not all possible combinations of displacements and rotations of the optical elements in a system will change the final position of the beam spot.  The derivation of the selection rules for the allowed terms is relatively straightforward.  The displacement of the beam spot in the detector ($x,y$) can be expressed to arbitrary order as a polynomial function of the displacements and rotations of the set of optical elements in the system
\begin{equation}
\label{poly}
(x,y) = f(\{dx\},\{dy\},\{dz\},\{d\alpha\},\{d\beta\},\{d\gamma\})
\end{equation}
where $\alpha$, $\beta$, and $\gamma$ are rotations about the $x$, $y$, and $z$ axes respectively and the curly braces represent the set of all perturbations of a given kind (i.e. \{$dx$\} represents the $x$ displacements of the set of optical elements).  If the entire system is reflected through the nominal axis of symmetry (or rotated about the axis of symmetry by $\pi$) then the beam spot must also reflect in a similar manner, $(x,y)\longrightarrow (-x,-y)$.  Therefore, each term in the polynomial function (\ref{poly}) must have odd symmetry in the plane transverse to the axis of symmetry.

From this discussion, one can define a preliminary, multiplicative, transverse parity operator $\pp$ such that transverse perturbations, whether displacements or rotations, change sign while the signs of longitudinal displacements or rotations remain unchanged
\begin{equation}
\label{parity}
\pp(x,y,z,\alpha,\beta,\gamma)\longrightarrow (-x,-y,z,-\alpha,-\beta,\gamma).
\end{equation}
Every possible perturbation to an optical element, whether a lens, detector, or mirror, etc. can be designated as having either even or odd transverse parity, though for a general optical element the transverse perturbations need not have odd transverse parity (hence ``preliminary'').  The expression for the final location of the beam spot can only be expressed using terms that have overall odd parity.

From the example of two lenses given earlier, each of the displacements of the lenses has odd transverse parity and the result (\ref{double}) has two terms, both linear in the transverse displacements and thus both with odd transverse parity.  A longitudinal displacement of a lens has even parity.  Such a displacement would allow a coupling of the form $dydz$ to contribute to the location of the beam spot since the overal parity of such a term is still odd
\begin{equation}
\pp(dydz) = (-dy)(dz) = -dydz.
\end{equation}
Each term in the rotation example also has odd parity.  The third term in equation (\ref{rotated}), for example, is proportional to $dyd\alpha^2$ which has parity
\begin{equation}
\pp(dyd\alpha^2) = (-dy)(-d\alpha)^2 = -dyd\alpha^2.
\end{equation}

One application of transverse parity is to shorten the calculation (to a given order) of the final location of the beam spot in a large optical system.  Instead of allowing the computer to calculate all the terms in the expansion, thereby spending a lot of time on terms that are ruled out because of incorrect parity, one can reject such terms immediately and focus on those that are allowed to contribute.

\subsection{Other Symmetries}

Other symmetries that are inherant in a particular optical element can further restrict the allowed terms in (\ref{poly}) and may affect the parity of a particular perturbation.  For example, a mirror has translational symmetry in the transverse plane and a cylindrical lens has translational symmetry along its axis of cylindrical symmetry.  A translation of a cylindrical lens along that axis or a rotation about that axis would produce no effect on the incident beam and would therefore have even transverse parity.

Symmetry arguments such as this also demonstrate why even-parity combinations of otherwise odd-parity perturbations are not allowed.  Consider the forbidden combination of $dyd\beta$ for a standard lens, a vertical translation followed by a rotation about the vertical axis.  Since the vertical translation maintains the left-right symmetry of a lens, no rotation about that same axis will produce a change in the location of the beam spot.

The parity of a given perturbation to a given optical element is ultimately defined by how that perturbation affects either the ray of light that passes through the geometric center of the optical element for translations or how it affects a ray of light that strikes the element at a point that is off of the rotation axis of a rotation-type perturbation.  A perturbation that produces a result with odd symmetry has odd transverse parity, and one that gives an even result would have even symmetry.  These definitions of parity only slightly change the parity operation given in (\ref{parity}) in that the odd parity perturbations will change sign and the even perturbations will not, regardless of the axis involved in the perturbation.

\subsection{A Real Beam}

Transverse parity can also describe the behavior of a beam with physical extent.  A ray that is displaced from the center of the beam would have an overall odd transverse parity.  The same is true for a ray that is not parallel to the longitudinal axis.  That transverse parity applies here is not surprising since, using a finite number of lenses and a central ray, one can reconstruct any other ray in the physical beam.  The physical beam is essentially a superposition of odd-parity, perturbed, central rays and each of the terms that are allowed in this construction must already have overall odd transverse parity.

\section{Discussion}

One example of the application of this study to an undergraduate course in optics would be the derivation of equations such as (\ref{single}) or (\ref{rotated}) using either a matrix optics approach or analytic geometry.  Another possibility, directly applicable to laboratory research, is to determine the sources of systematic error in a given optical system and their relative importance.  Such a study could include how those errors scale with the nominal placement of the optical elements (e.g. the separation of a light source and a collomating lens) or with perturbations to the elements.  The introduction of a cylindrical lens presents an opportunity to study the translational symmetry of the lens and also the effects that arise from the addition of the ``preferred direction'' along the axis of cylindrical symmetry.

Beyond the direct application to optics, there are many aspects of this work that pertain to many branches of physics.  Selection rules, parity, symmetry, dimensionless ratios, and perturbations are ubiquitous in physics and can all be studied with this model.  A student could be asked to determine if a given coupling is allowed, to determine the coefficient that corresponds to that perturbation, or even to empirically find the analytic expression for that same coefficient by using the dimensionless ratios that can be formed by the separation of the lenses and their focal lengths.  Similarly, a student could study how the coefficients scale with the nominal separation of the optical elements, such as the expression that multiplies the perturbation to the first lens in equation (\ref{double}) which scales linearly with short separations and quadratically with large separations.

Using the perturbed optical system one can learn about many important tools that physicists use in their efforts to understand natural law.  The apparatus, consisting of a beam of light and a few lenses, is relatively inexpensive and easy to manipulate.  The effects of a perturbation to the system appear instantly on a screen or detector which transparancy allows for a deeper investigation of the physical principles being studied instead of simply a tutorial on the use of a particular piece of laboratory equipment.  This system should prove useful as a pedagogical device for undergraduates, both as a theoretical tool and as an experimental investigation of a simple theory.

\begin{acknowledgements}
I would like to thank Drs. Michael Moore and Paul Boynton for their comments pertaining to this work.
\end{acknowledgements}

\end{document}